# A Hard-Science Approach to Kondratieff's Economic Cycle

Theodore Modis*

* Theodore Modis is the founder of Growth Dynamics, an organization specializing in strategic forecasting and management consulting: http://www.growth-dynamics.com
Address reprint requests to: Theodore Modis, Via Selva 8, 6900 Massagno, Lugano, Switzerland. Tel. 41-91-9212054, E-mail: tmodis@yahoo.com

## Highlights

- Kondratieff's economic cycle (K-cycle) can be evidenced via variations in the rate of energy consumption.
- Several other human endeavors/phenomena resonate with the K-cycle.
- The K-cycle may have its origins in a climatic variation or in the active human lifespan.
- All evidence for its existence carries confidence levels that are poor by scientific standards.
- There is some evidence that the K-cycle may be beginning to wash out.

## Abstract

In an effort to evidence the Kondratieff cycle more scientifically than the way economists do, physical variables are studied rather than monetary indicators. Previously published graphs are reproduced and updated here with recent data. A cyclical rather regular variation of energy consumption reveals a 56-year cycle. A dozen human endeavors/phenomena, such as bank failures, homicides, hurricanes, feminism, and sunspot activity are shown to resonate with this cycle. Possible explanations for this phenomenon may have to do with a climatic variation or with the length of time any individual actively influences the environment. There is some evidence that the cycle may be getting shorter in amplitude and duration in recent years. All quantitative confidence levels involved in these observations are poor by scientific standards and permit critics to question the very existence of this phenomenon.

## 1. Introduction

Claims for long waves in economic activity have existed since the beginning of the Industrial Revolution. Among the early proponents of economic cycles was William S. Jevons (1835-1882) who linked business cycles to sunspot activity.[1] Later Henry Ludwell Moore (1869-1958) linked business cycles to climate variations arguing that a rainfall cycle affects agricultural markets, which affect industrial markets.[2] The Russian economist Nikolai D. Kondratieff (1892-1938) deduced an economic cycle with a period of about fifty years from economic indicators alone. His classic work in 1926 resulted in his name being associated with this phenomenon.[3] Joseph A. Schumpeter (1883-1950) tried to explain the existence of economic cycles and in particular Kondratieff's cycle by attributing growth to the fact that major technological innovations come in clusters.[4] More recently, Bert de Groot and Philip Hans Franses have found a multiplicity of cycles in innovations.[5] And Andrey V. Korotayev, Julia Zinkina, and Justislav Bogevolnov have evidenced Kondratieff waves in global invention activity.[6]

One could argue that Kondratieff's cycle is the most successful among long-wave postulations. His name yields a quarter of a million hits in a Google search, and an economic research organization called International N. D. Kondratieff Foundation has been established in 1992 accredited by the Russian Academy of Sciences. Its charter is to coordinate interdisciplinary research, organize conferences and competitions, and award medals to Kondratieff-related contributors. In Russian economic circles the whole thing takes on the airs of a cult with an inexhaustible list of publications, see for example issues of the *Kondratieff Waves* yearbook.[7]

And yet, Kondratieff's work has been challenged by many respected economists from the very beginning. Critics doubted both the existence of Kondratieff's cycle and the causal explanation suggested by Schumpeter. Among vocal critics has been American economist Murray Rothbard.[8] He argued that business cycles are "emphatically not periodic." He called the Kondratieff cycle "mystical" and "the flimsiest 'cycle' of them all." He questioned and discounted Kondratieff booms/depressions, and presented arguments showing that the Kondratieff cycle may seem regular at the very most for only three-and-a-half periods. He also criticized the fact that it is evidenced by studying prices, which do not accurately reflect the state of the economy.

Kondratieff's postulation ended up being largely ignored by contemporary economists for a variety of reasons. Since then it came in and out of vogue with changes in the economic climate. In the final analysis, however, the postulation's greatest weakness may have been the boldness of the conclusions drawn from such ambiguous and imprecise data as monetary and financial indicators. These indicators—just like price tags—are a rather frivolous means of assigning lasting value. Inflation and currency fluctuations due to speculation or politico-economic circumstances can have a large unpredictable effect on monetary indicators. Extreme swings have been observed. For example, Van Gogh died poor, although each of his paintings is worth a fortune today. The amount of work or beauty in his paintings has not changed since his death; counted in dollars, however, it has increased tremendously. Even the monitoring of innovations and invention activity is subject to human bias and uncertainty that stem from the ambiguity involved in defining them and quantifying them.

A number of “hard” scientists have attempted to evidence Kondratieff’s cycle by studying “physical” variables such as homicides and energy consumption. Deaths and watts consumed are not subject to speculation; they are unambiguously defined, and precisely measured. It was Hugh B. Stewart, a physicist, who first studied cyclical variations in energy consumption in America. He extracted a rather regular cyclical variation above and below the long-term trend of energy consumption in the U.S.[9] Cesare Marchetti, another physicist, replicated that cyclical variation, including more recent data, and demonstrated with a fair amount of success that many other social endeavors are synchronized with it.[10] I replicated and augmented Marchetti’s work in my book *Predictions: Society’s Telltale Signature Reveals the Past and Forecasts the Future*.[11]

In the following sections I will reproduce some of the most convincing evidence for the existence of Kondratieff’s cycle using physical variables and three more decades of data.

## 2. Energy Consumption

There are historical data on energy consumption in the U.S. going back to 1850.[12] In Figure 1 we see the evolution of this variable up to the end of 2015 plotted versus time with 5-year sampling. The growth seems to be stepwise with two long steps and a shorter recent one. An logistic S-curve fitted to the entire range via a Chi-square minimization does a mediocre job describing the overall growth pattern. Smaller S-curves describe better the three growth steps. Similar graphs with only the first two steps have been previously published.[13] At that time the third step had been sketched in—with an S-curve similar to the previous two—as a probable scenario for the future.

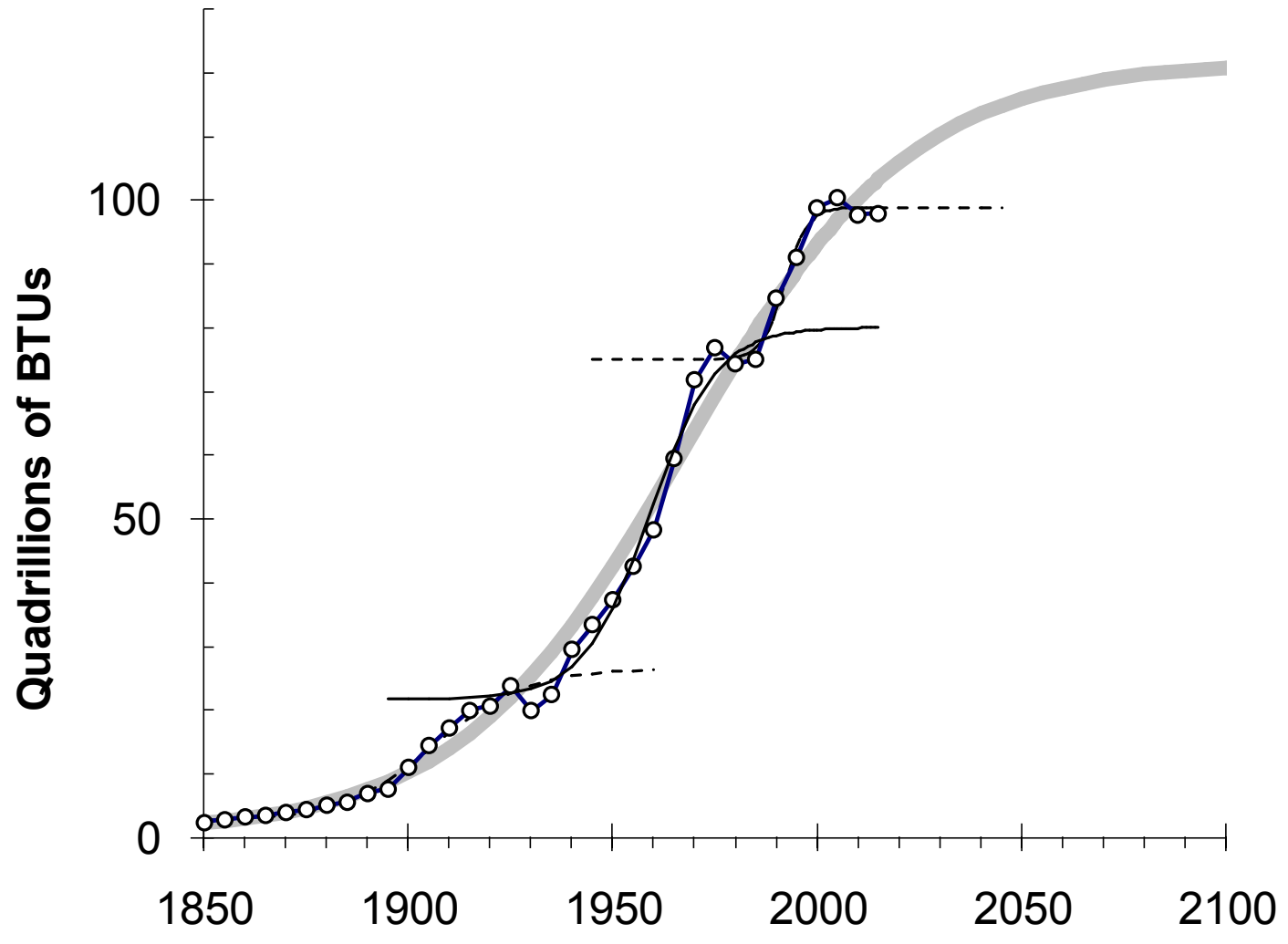

**Figure 1** Annual energy consumed in the U.S. sampled every five years. An overall logistic curve (thick gray line) fitted on the data helps identify three smaller sub-processes more amenable to logistic fits (thin and intermittent lines.)

Data sources: *Historical Statistics of the United States, Colonial Times to 1970*, vol. 2, Bureau of the Census, Washington, DC. Recent data from the BP *Statistical Review of World Energy 2016*.

In Figure 2 the deviations of the data from the overall S-curve trend have been isolated by taking the ratio of data to trend in Figure 1. The consumption of electrical energy in the U.S. treated in the same fashion has been superimposed on the same graph. A sinusoidal wave with period 56 years—thick gray band—is there to guide the eye through a regular oscillation.

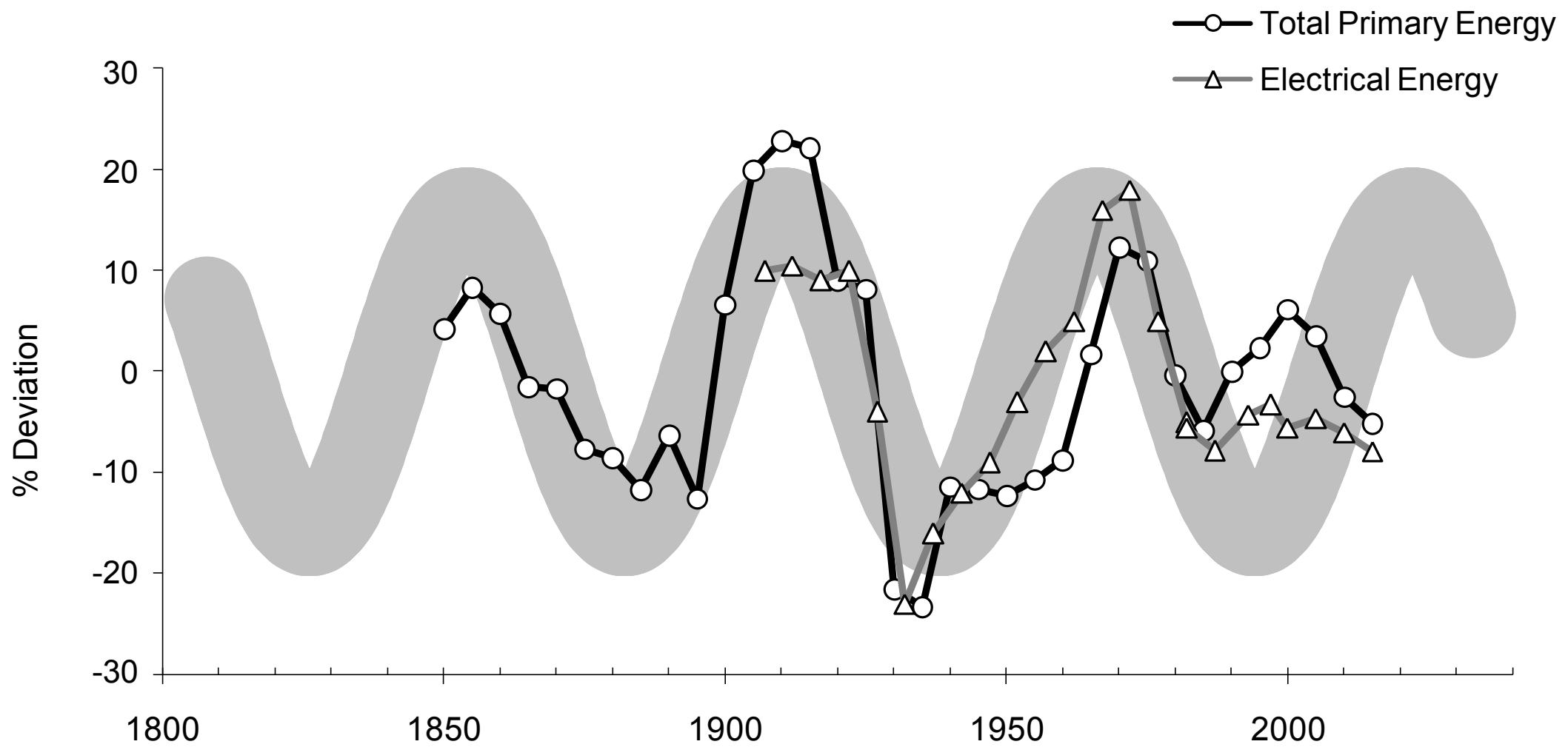

**Figure 2** Deviations from a logistic-growth trend for annual energy consumption in the U.S. (see Figure 1). Total primary energy (little circles); electrical energy (little triangles). The wide gray band is a sine wave with period 56 years. The correlation between total energy and sine wave is **r** = 0.63 and between electrical energy and sine wave **r** = 0.79.
Data sources for electrical energy: *Statistical Abstract of the United States*, U.S. Department of Commerce, Bureau of the Census. Recent data from the U.S. Energy Information Administration (EIA) *Monthly Energy Review*, January 2017.

A similar approach can be applied to data concerning worldwide energy consumption per capita. Ausubel et. al. have published a graph similar to that of Figure 1 with data up to 1985.[14] I have updated that graph to the end of 2015 with data from the BP *Statistical Review of World Energy 2016*, and the Bureau of the Census, Washington, DC. Isolating deviations of the data from an overall S-curve trend, as was done earlier, I obtained the graph in Figure 3.

There is an evident rather regular oscillation of the data with period of 56 years in both Figures 2 and 3. It seems that energy is consumed more ravenously at some times than at others. Whether we look at the U.S. or at the entire world, energy consumption has been as much as 20% higher than we would have expected during some periods, and as much as 20% lower than we would have expected in other periods. Enhanced energy consumption translates to enhanced economic growth and prosperity whereas diminished energy consumption reflects economic recession, stagnation or depression. In other words, Figures 2 and 3 produce independent evidence for an economic cycle with a period of 56 years otherwise known as Kondratieff's economic cycle.

It is noteworthy that there is a phase difference between the regular waves of Figures 2 and 3, namely booms and busts come seven years earlier in the U.S. than worldwide. Another observation is that the U.S. data deviate significantly—both in timing and in amplitude—from the regular cyclical pattern beginning in 1990.

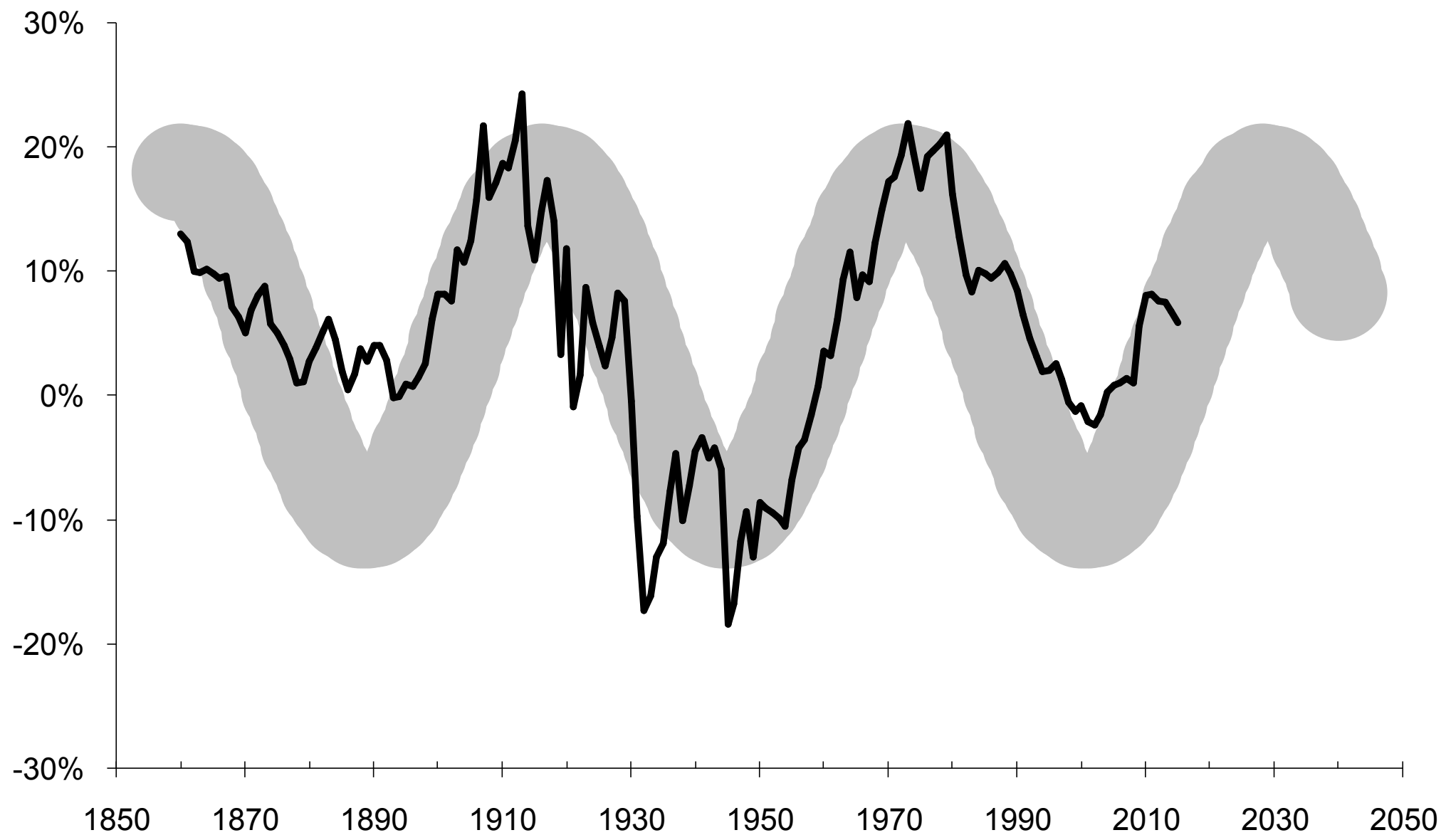

**Figure 3** Deviations from an S-curve trend on world energy consumption per capita. Annual data for the total energy per capita consumed worldwide (black line). The wide gray band is a sine wave with period 56 years. The correlation between total energy and sine wave is **r** = 0.72.

The correlation coefficient **r** between the time-series data and the idealized sine wave is perhaps more useful when expressed as $\mathbf{r}^2$ because it then represents the amount of structure in the data pattern that can be explained in terms of the regular sine-wave pattern. For the three variables plotted in Figures 2 and 3, namely total U.S. energy, electrical U.S. energy, and worldwide energy per capita we have respectively 39%, 62%, and 51% of their pattern explained by the regular sine wave shown.

## 3. Other Phenomena Resonating with Kondratieff's Cycle

In this section I will reproduce and update some other physical variables that have been seen to resonate with Kondratieff's cycle. In each figure a wide gray band representing a regular sine wave is sketched in to guide the eye.

*Bank Failures*

Figure 4 shows bank failures in the U.S., bank suspensions before 1933, and banks closed due to financial difficulties between 1933-2013. It is not surprising that bank failures peak close to the troughs of the energy-consumption cycle.

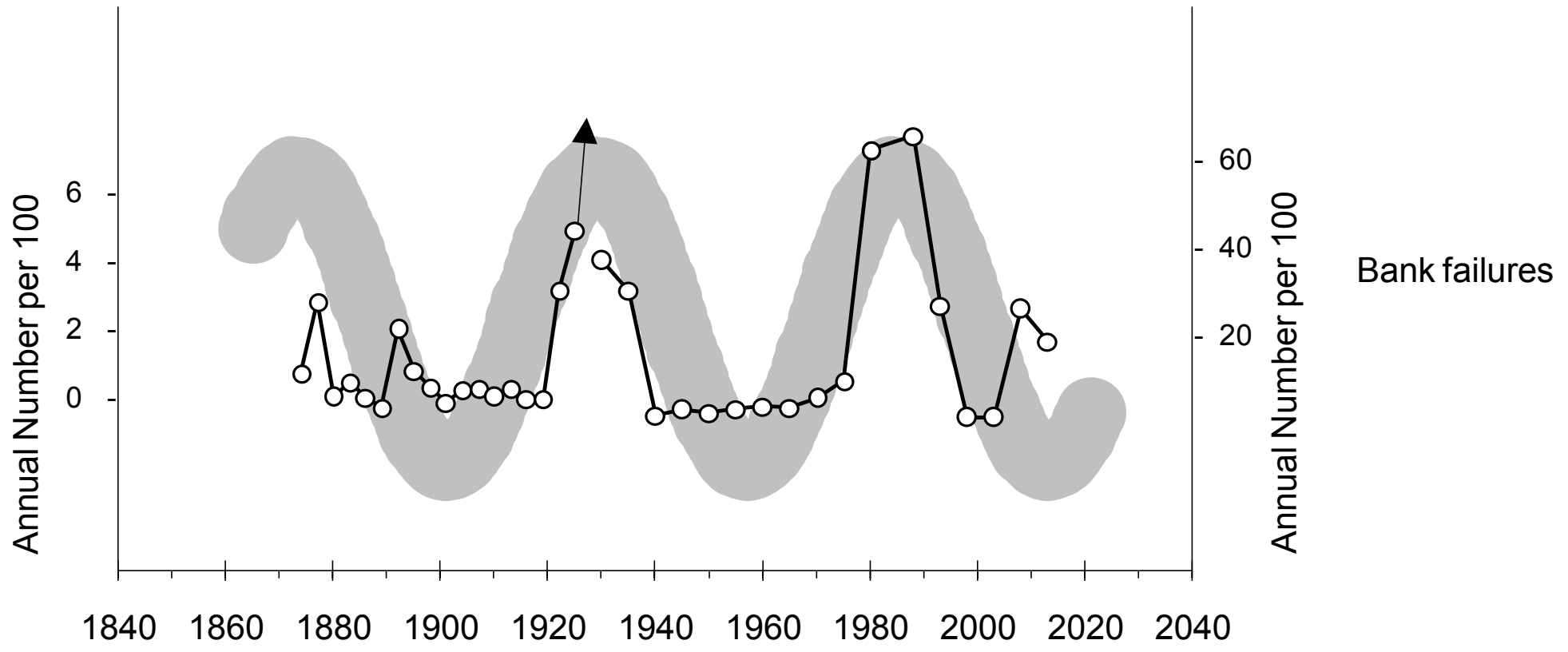

**Figure 4** Bank failures in the U.S. The small circles show data with 3-year sampling before 1933 (left vertical axis) and 5-year sampling after 1933 (right vertical axis). The gray band is a sine wave with period 56 years. Correlation **r** = 0.55, $\mathbf{r}^2$ = 31%.
Data sources: *Statistical Abstract of the United States*, U.S. Department of Commerce, Bureau of the Census. After 1933, Federal Deposit Insurance Corporation.

*Innovations and Discoveries*

At the top of Figure 5 we see the appearance of basic innovation in 10-year time bins as they are defined by Gerhard Mensch.[15] The exact number of innovations may be subject to debate and personal bias cannot be excluded in their definition. The graph has been updated after 1960 with data on the number of patents for inventions worldwide, which has been taken as a proxy for the appearance of innovations for that period. On the right-hand vertical axis we see the percent deviation from a logistic-growth trend fitted on the total number of patents.

The variation over time for both the number of innovations and the deviations from the patent trend seem well synchronized with a cycle of 56 years (gray band). The peaks line up with the troughs of the energy cycle of Figure 1 (the Kondratieff cycle). One could understand why innovations increase during economic hardship. It follows from the natural reaction of people to become more entrepreneurial when economically squeezed. But this reasoning conflicts with Schumpeter's explanation for the existence of Kondratieff 'cycle namely that it is caused by the clustering of innovations. What comes first the clustering of innovations or the Kondratieff'cycle? What is the cause and what the effect?

A more objectively and precisely defined variable is the number of basic chemical elements discovered per decade, shown at the bottom of Figure 5.[16] Again there seems to be resonance with a regular cycle of 56 years (gray band) identical to that of the innovations graph. One may be tempted to establish a causal relationship; the technologies for discovering and separating elements could themselves be linked to the basic innovations and therefore would display the same bunching pattern. But despite the fact that four out of five peaks in element discovery line up well with the sine-wave peaks, the correlation coefficient turns out to be only **r** = 0.22 ($\mathbf{r}^2$ = 5%) implying that only 5% of what we see on the elements pattern can be explained by the regular sine-wave patter.

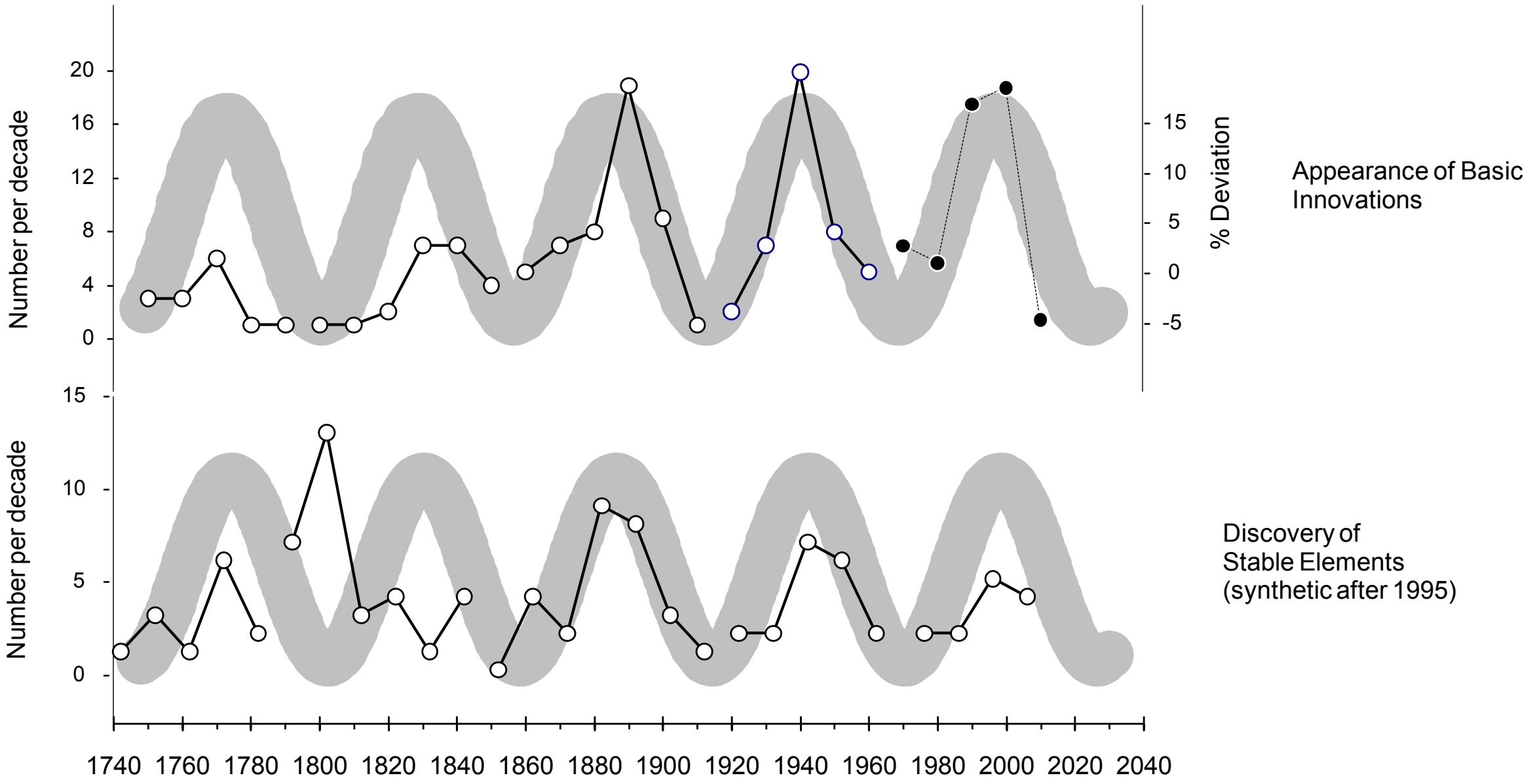

**Figure 5** Basic innovations and stable elements. At the top we see the appearance of innovations (open circles) on the left vertical axis; also deviations on a trend for the appearance of patents (black dots) on the right vertical axis. At the bottom the discovery of the stable elements; the last group concerns artificially created elements. 10-year time bins everywhere. The gray bands are identical sine waves with period 56 years.
Innovations: Correlation **r** = 0.59, $\mathbf{r}^2$ = 35%
Elements: Correlation **r** = 0.22, $\mathbf{r}^2$ = 5%
Data sources: For innovations Gerhard Mensch. For patents World Intellectual Property Organization (WIPO). For stable elements *The American Institute of Physics Handbook*, 3rd ed. (New York: McGraw-Hill) and Wikipedia "Timeline of chemical element discoveries".

*Cirrhosis Victims*

In Figure 6 we see the annual death rate of victims from cirrhosis of the liver sampled every 10 years until 1980, and every 5 years afterward. I have published such a graph in 1992 with data up to 1985.[11] There is no overall trend but a rather pronounced cyclical variation. Once again there is good synchronicity with a cycle of 56 years (gray band). The number of victims reaches a maximum during periods corresponding to boom years in the Kondratieff cycle of Figure 1, as if enhanced prosperity accentuates this illness whose main cause is alcoholism.

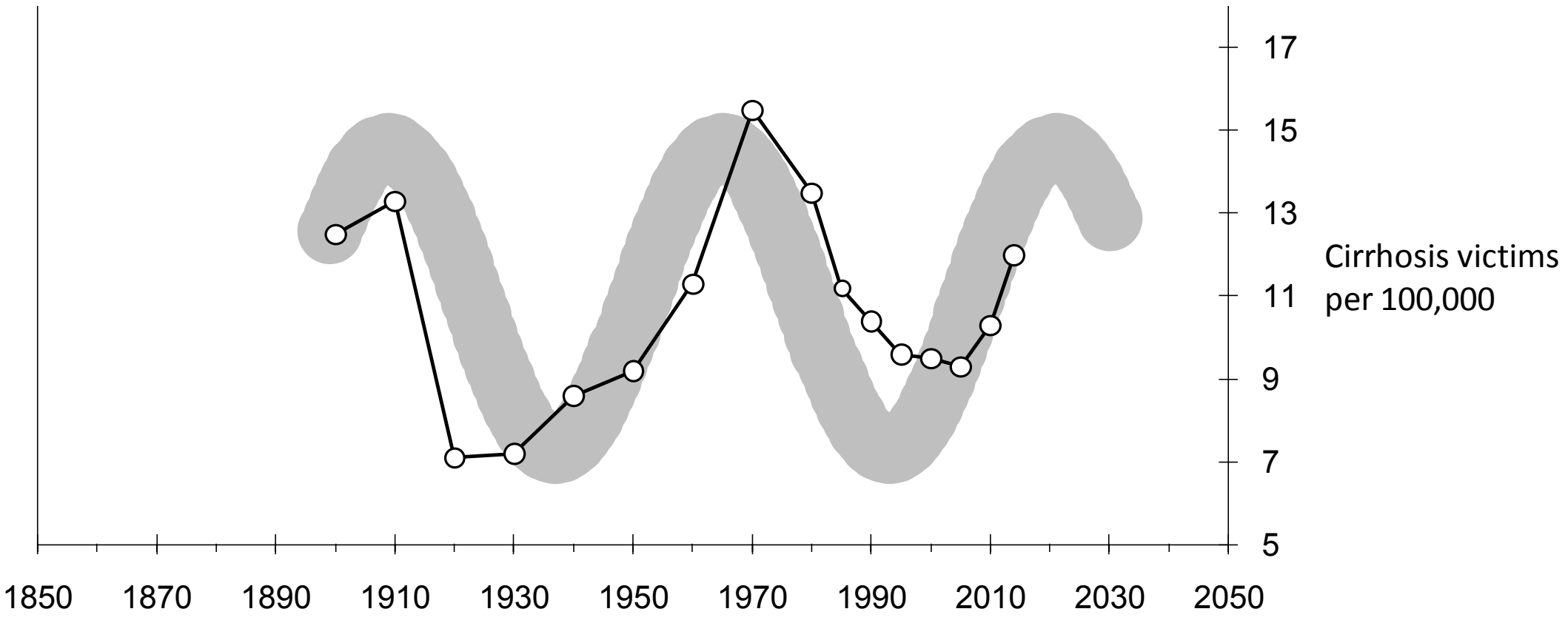

**Figure 6** Cirrhosis of the liver in the U.S. Deaths per 100,000 population (open circles). The gray band is a sine wave with period 56 years. Correlation **r** = 0.58, $\mathbf{r}^2$ = 33%
Data source: *Statistical Abstract of the United States*, U.S. Department of Commerce, Bureau of the Census. Recent data from the National Center for Health Statistics, Chronic liver Disease and Cirrhosis, 2014.

*Homicides*

Homicides in the U.S. also rise and fall periodically with no particular trend. The phenomenon was pointed out by Marchetti in 1986 with data up to 1975.[10] Figure 7 shows the evolution of homicides per 100,000 population with 5-year sampling updated to 2014. The correlation with a regular sine wave—of a somewhat shorter period, 54 year—is the best one yet, **r** = 0.87. Homicides peak during the declining phases of the Kondratieff cycle.

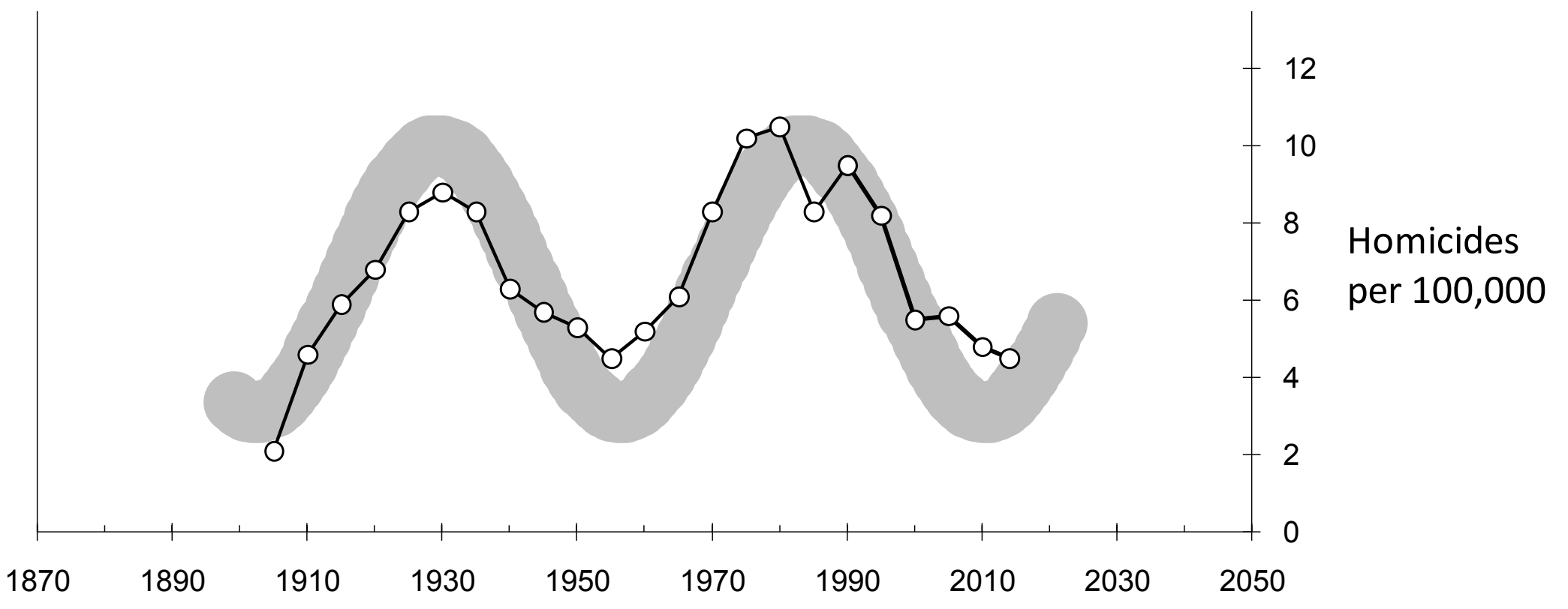

**Figure 7** Homicides per 100,000 population in the U.S. The gray band is a sine wave of period 54 years. Correlation **r** = 0.87, $\mathbf{r}^2$ = 75%.
<u>Data source:</u> *Statistical Abstract of the United States*, U.S. Department of Commerce, Bureau of the Census. After 1990 from FBI, Crime in the United States 2014.

*Feminism*

The percentage of women among Nobel laureates has been used as a proxy—albeit an arbitrary one—for feminism with data up to 1980.[11] Figure 8 shows this percentage of women since the beginning of Nobel prizes and up to 2010 in 10-year time bins. A variation that was in good agreement with a 54-year sine wave (thick gray band) seems to break down in 2010.

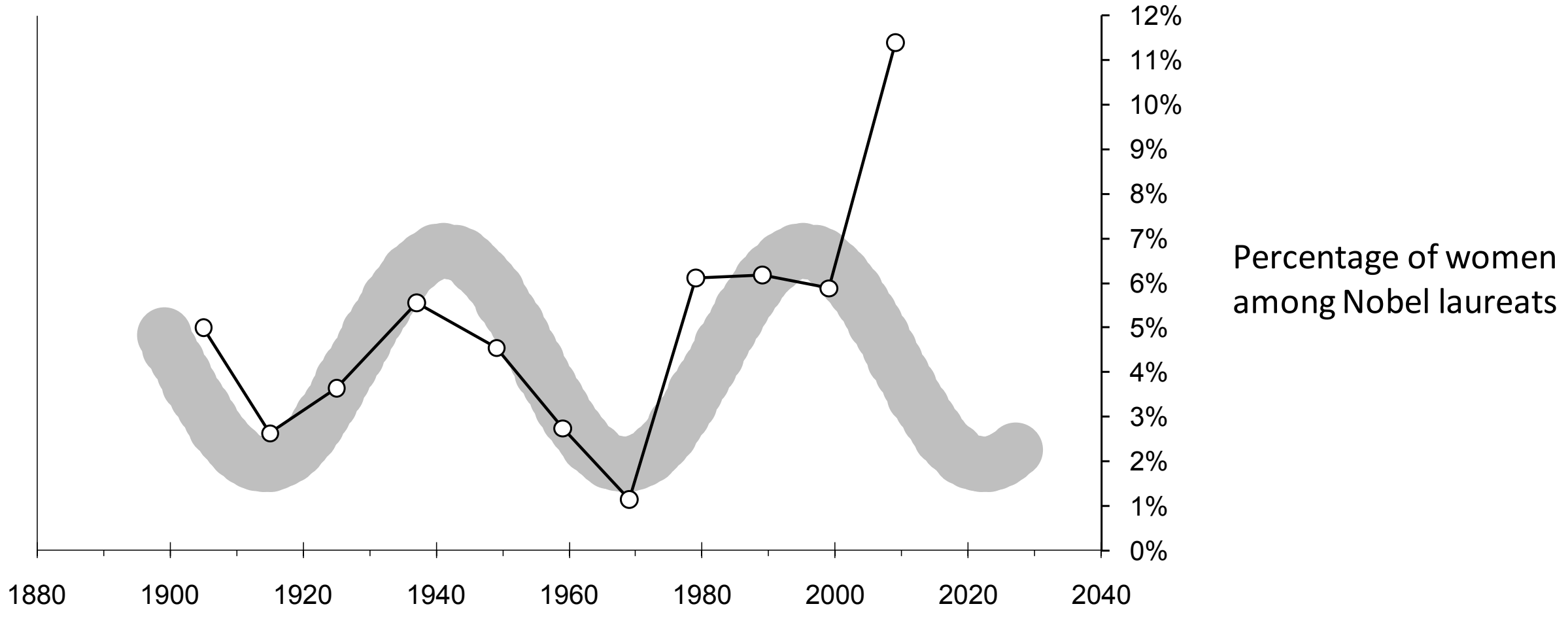

**Figure 8** The percentage of women among Nobel laureates in 10-year time bins. The regular sine-wave gray band has a period of 54 years. Correlation **r** = 0.47, $\mathbf{r}^2$ = 22%.
<u>Data source:</u> Nobel Foundation.

*Hurricanes*

The evolution of the number of hurricanes (category ≥ 3) over the Atlantic has been studied since 1851.[17] Figure 9 reproduces here the percentage deviation of the number of hurricanes per decade from a gently rising overall S-curve trend. A time series with 10-year time bins shows a rather regular variation in good agreement with a 56-year sine wave (thick gray band).

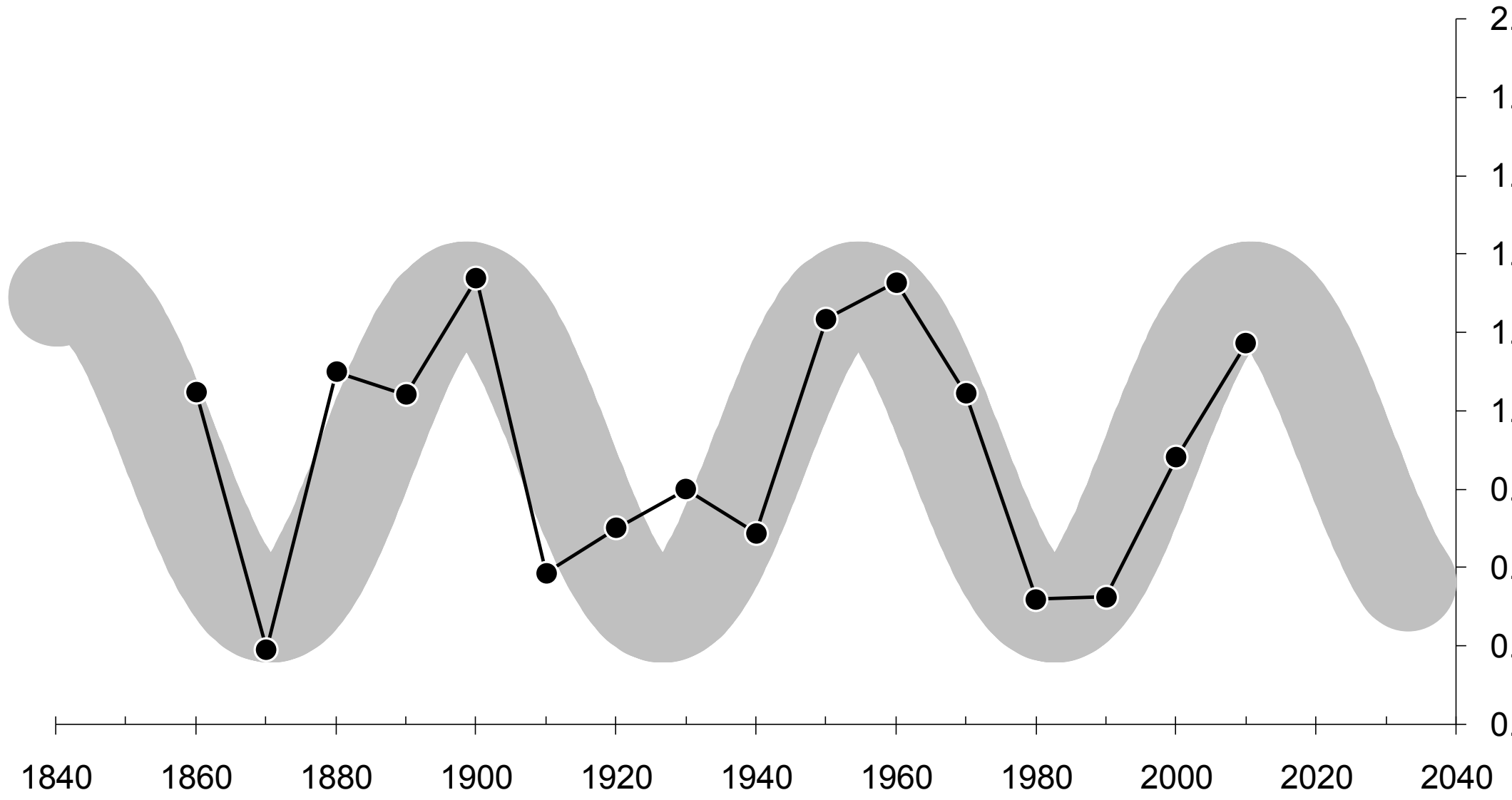

**Figure 9** Atlantic Hurricanes. Ratio of the data to an S-curve trend. The thick gray band is a sinusoidal wave with a period of 56 years. Correlation **r** = 0.75, $\mathbf{r}^2$ = 56%.
Data source: Unisys Weather, Hurricane/Tropical Data (data courtesy of Tropical Prediction Center) http://weather.unisys.com/hurricane/

*Sunspot Activity*

There is a well known regular 11-year variation in sunspot intensity. During maximum sunspot activity, the overall solar output increases by a few tenths of 1 percent. The corresponding temperature change on the earth may be too small to be felt, but meteorologists in the National Climate Analysis Center have incorporated the solar cycle into their computer algorithms for the monthly and the 90-day seasonal forecasts.

In the 300-year-long history of documented sunspot activity, we can detect relative peaks in the number of sunspots every fifth period (5 x 11 = 55 years). To do this we must extract the variation in the number of sunspots smoothed over a rolling 23-year period with respect to a 54-year moving average. Figure 10 shows the results of such a procedure—routinely used in time series analyses—that washes out the well-known 11-year cycle of sunspot activity and reveals a longer periodic variation similar to the energy-consumption cycle. With one exception—a peak missing around 1900—the oscillation shows a fair agreement with Kondratieff's cycle. This result was first observed with data up to 1985 smoothing over a rolling 20-year period with respect to a 56-year moving average.[11]

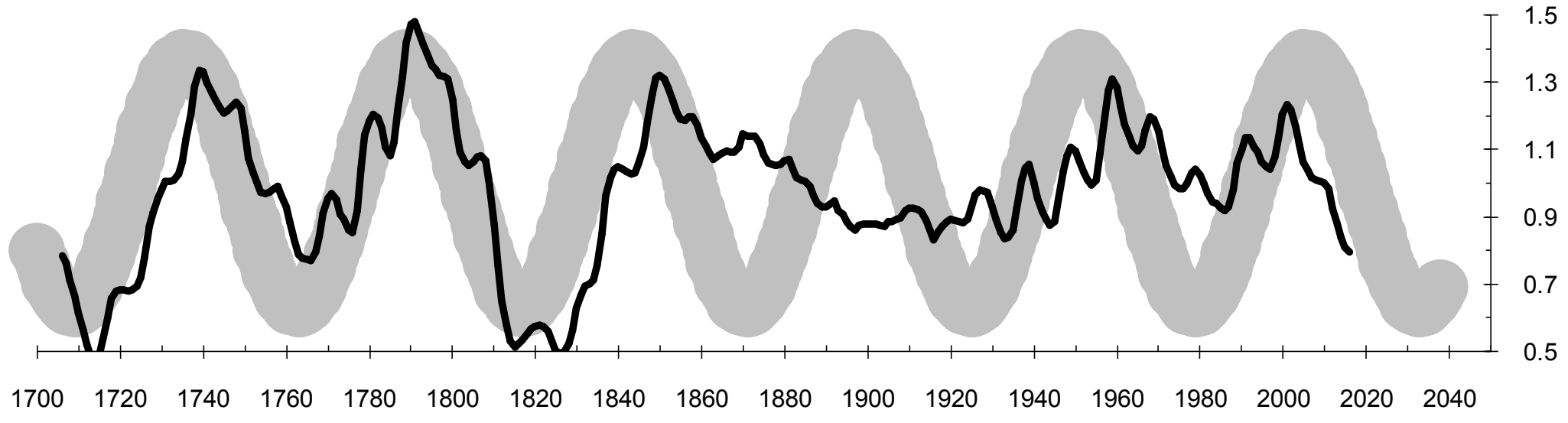

**Figure 10** Long-term variation of sunspot activity. Yearly data on sunspot activity smoothed over a rolling 23-year period with respect to a 54-year moving average. The gray band is a sine wave with period 54 years.
Correlation $\mathbf{r} = 0.45$, $\mathbf{r}^2 = 20\%$.

*Price Index*

For the sake of completeness I will also consider a non-physical variable, the wholesale price index in the U.K. This variable has been traditionally used to evidence the Kondratieff cycle because of its long historical record.[18] Figure 11 shows the U.K. Wholesale Price Index updated to 2014 and smoothed over a rolling 27-year period with respect to a 55-year moving average. This procedure washes out small fluctuations and reveals a wave. The little stars point out peaks and valleys. The average periodicity turns out to be 55.8 years. Visually there is convincing evidence for a correlation between the price index and a regular sine wave of period 55.8 years. However, the calculated correlation is rather poor; only 8.3% of the index pattern can be explained by the sine wave.

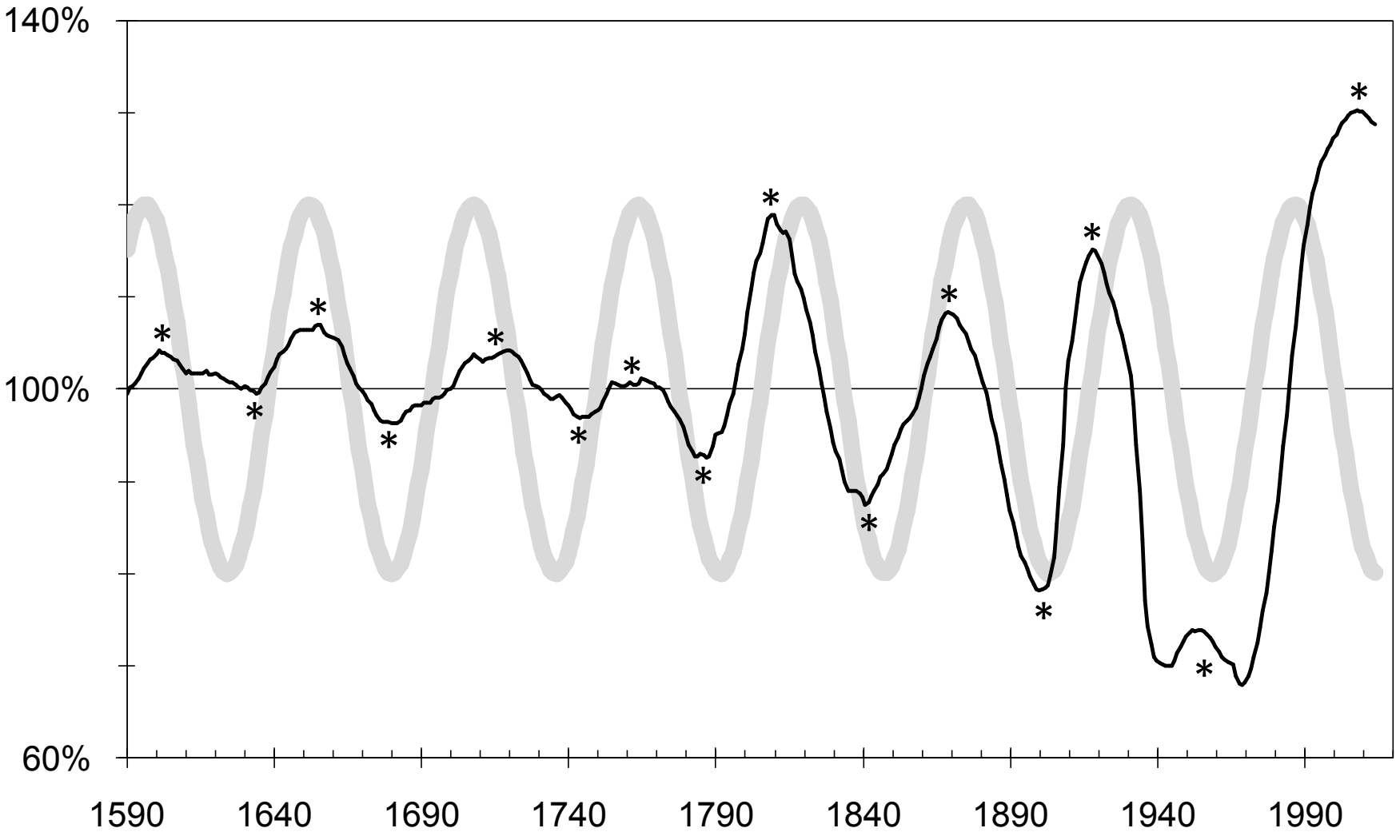

**Figure 11** The U.K. Wholesale Price Index smoothed over a rolling 27-year period with respect to a 55-year moving average. The little stars point out peaks and valleys. The gray line is a sine wave with period 55.8 years. Correlation **r** = 0.29, $\mathbf{r}^2$ = 8.3%

## 4. Possible Explanations for the Kondratieff Cycle

In *Predictions* (1992) I have suggested two possible explanations for Kondratieff's cycle independent of those made by economists. One is linked to periodic phenomena in celestial motions. There are cosmic events whose influence on the earth originate far from our planet and pulsate with a beat of about 56 years. The cycle of Saros, known since antiquity, is based on the fact that identical solar and lunar eclipses occur every eighteen years, eleven days, and eight hours but will not be visible at the same place on the earth. The cycle of Meton, which has been used in the calculation of the date of Easter, is based on the fact that every nineteen years the same lunar phases will occur at approximately the same time of the year. In fact, lunar eclipses recur and are visible at the same place on the earth every 18.61 years. Therefore, the smallest integral year time unit that allows accurate prediction of eclipses at the same place is nineteen plus nineteen plus eighteen, a total of fifty-six years. This fact played an important role in the construction of Stonehenge where there are 56 carefully spaced and deeply dug holes, the so-called Aubrey holes. By using them to count the years, the Stonehenge priests could have kept accurate track of the moon, and so have predicted danger periods for the most spectacular eclipses of the moon and the sun. In fact, the Aubrey circle could have been used to predict many celestial events.[19]

Lunar and solar eclipses figure prominently in superstition, but their importance goes beyond that. Biological effects, including strange animal behavior, have been observed during eclipses. The mere effect of having the three celestial bodies on a straight line provokes exceptional tides. There are periodicities of 56 years on the prediction of tides.[20] But the 56-year period concerns not only eclipses and the alignment of the earth, moon, and sun on a straight line. Any configuration of these three bodies will be repeated identically every 56 years. Possible effects on the earth linked to a particular geometrical configuration will vary with the 56-year period. Marchetti has told me that continental shelves and rings in the trunks of many-century-old trees have shown structures pointing at such a periodicity. The implication is that the climate may be changing periodically. The variations seen in the number of hurricanes and sunspot activity from Figures 9 and 10 respectively corroborate a possible link to the climate.

Another possible explanation for the Kondratieff cycle was suggested to me by Nobel-laureate physicist Simon van der Meer while we were discussing these observations and the fact that a period of 56 years is close to the length of time any individual actively influences the environment. Being an expert in stochastic processes, van der Meer suggested that individuals could be acting as fixed "delays" in a never-ending flux of change. In society there are many feedback loops, and despite a continuous arrival of individuals, the existence of a fixed delay—triggered by some instability—could produce "bunching" in social phenomena with a characteristic period equal to the delay.

Whether it is the climate or man's active lifespan behind it, there could be reason to see a modulation of the economy or other human endeavor with a cyclical variation of about 55 years. Correlated activities would then also be impacted.

## 5. Discussion

There are some deviations of the data from the regular cyclical pattern. The cycle of energy consumption shown in Figure 2 seems to diminish in amplitude and shorten in period toward the end of the 20$^{th}$ century and the beginning if the 21$^{st}$. Given that the U.S. leads the world by seven years or so, conceivably these irregularities will soon show up in the worldwide data of Figure 3. Other irregularities during recent times like those seen for bank failures and women Nobel laureates could be related. Could it be that the Kondratieff cycle is diminishing as a phenomenon? A recent article by Russian economists raises the possibility of the end of this phenomenon based on globalization arguments.[21]

Another reason for a diminishing Kondratieff cycle may be reflexivity, namely the fact that once people become aware of a situation, their behavior may try to change the situation to their advantage. This notion—akin to the notion of a self-fulfilling prophecy and the Efficient Market Hypothesis (EMH)—has been known in one form or another throughout the 20$^{th}$ century, be it in sociology or in the world of investments as EMH. Reflexivity received more limelight in economics following the crash of 2008.

Despite the irregularities mentioned above I believe that Figures 2-11 present a compelling case—at least visually—for the existence of Kondratieff's cycle. And dealing with unambiguously defined and accurately measured *physical* variables enhances the reliability of this conclusion. In the wake of the publication of my first book *Predictions*—in which most of these graphs were presented with data that stopped in the late 1980s—a reviewer in *Science* wrote:[22]

> "... he (Modis) does cite the relationship of his 'overall' 56-year cycles and evidence on the associated clustering of technological innovations to prior work by Kondratieff and Schumpeter (in my opinion, Modis' evidence on these topics is more extensive and compelling than that of either of these scholars.)"

And yet, the average correlation from all the graphs in these ten figures is 0.576 implying only 33% of the patterns' behaviors can be explained by a regular cycle (a sine wave).

In my career I have published over 100 articles in scientific and business journals, some of which I am proud of. Unfortunately, the publication that has drawn most attention, if I judge by the number of reads reported in ResearchGate, is an article titled "Sunspots, GDP, and the Stock Market" that reports evidence for *some* correlation between stock-market moves and sunspot activity.[23] However, it is possible that a series of numbers generated randomly also correlates in a non-negligible way with a given time series.

I have made a simulation study to explore the possible correlation of randomly generated data with sunspot-activity data. I generated 25,000 time series with random data normally distributed with an average and a sigma equal to those of the sunspot data. Figure 12 shows one example where the random data seem to correlate significantly with the sunspot data. For every one of the sunspot peaks (intermittent line), rather regularly spaced every 11 years, there is a corresponding peak in the random data (gray line), and six of the eight peaks line up rather well. An uninformed reader would be hard-pressed to say that all this lining-up is accidental. And yet, the correlation coefficient is 0.42 implying that only 18% of the gray line can be explained by the intermittent line.

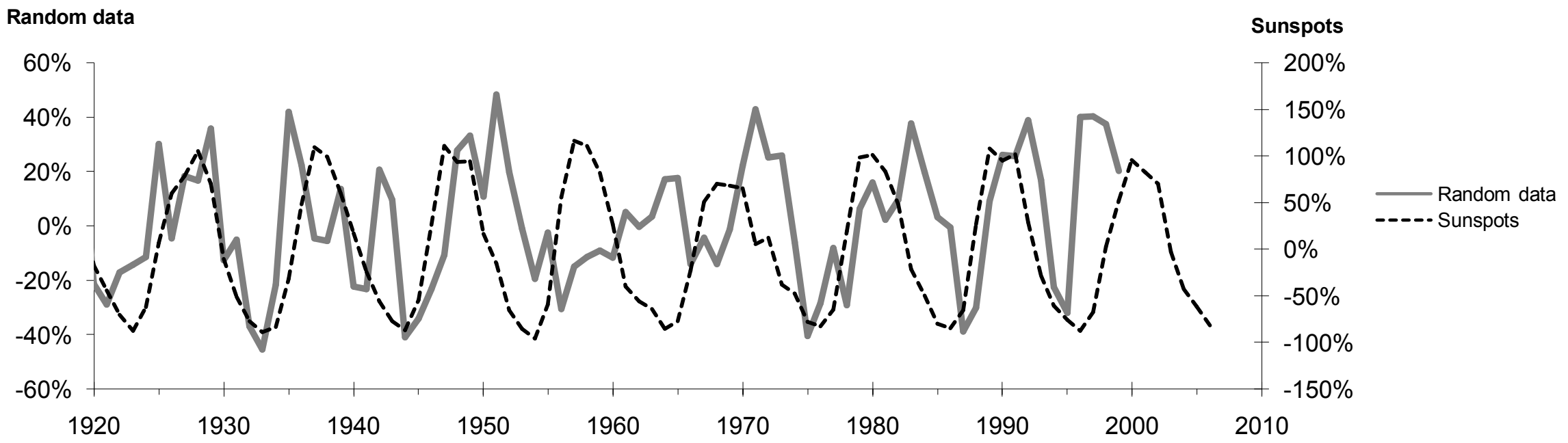

**Figure 12** Random Data and Sunspots. Sunspot activity (intermittent line) and randomly generated data (gray line). Correlation **r** = 0.42, $\mathbf{r}^2$ = 18%

Figure 13 shows a histogram of the correlations with sunspots for the 25,000 randomly generated events. As expected, the average correlation is zero, meaning there is no correlation. But there is a distribution indicating a probability of 2.5% for correlations with $|\mathbf{r}| \geqq 0.45$. In other words 25 times in a thousand, such correlation—which is not far from the average correlation in Figures 2-11—can emerge with randomly generated data. And this should be taken as a limit because similar simulations for variables with less structure than sunspots—Figures 6-8 display only 2 to 3 periods as opposed to 8 periods for sunspots—would yield more correlation with random events more easily.

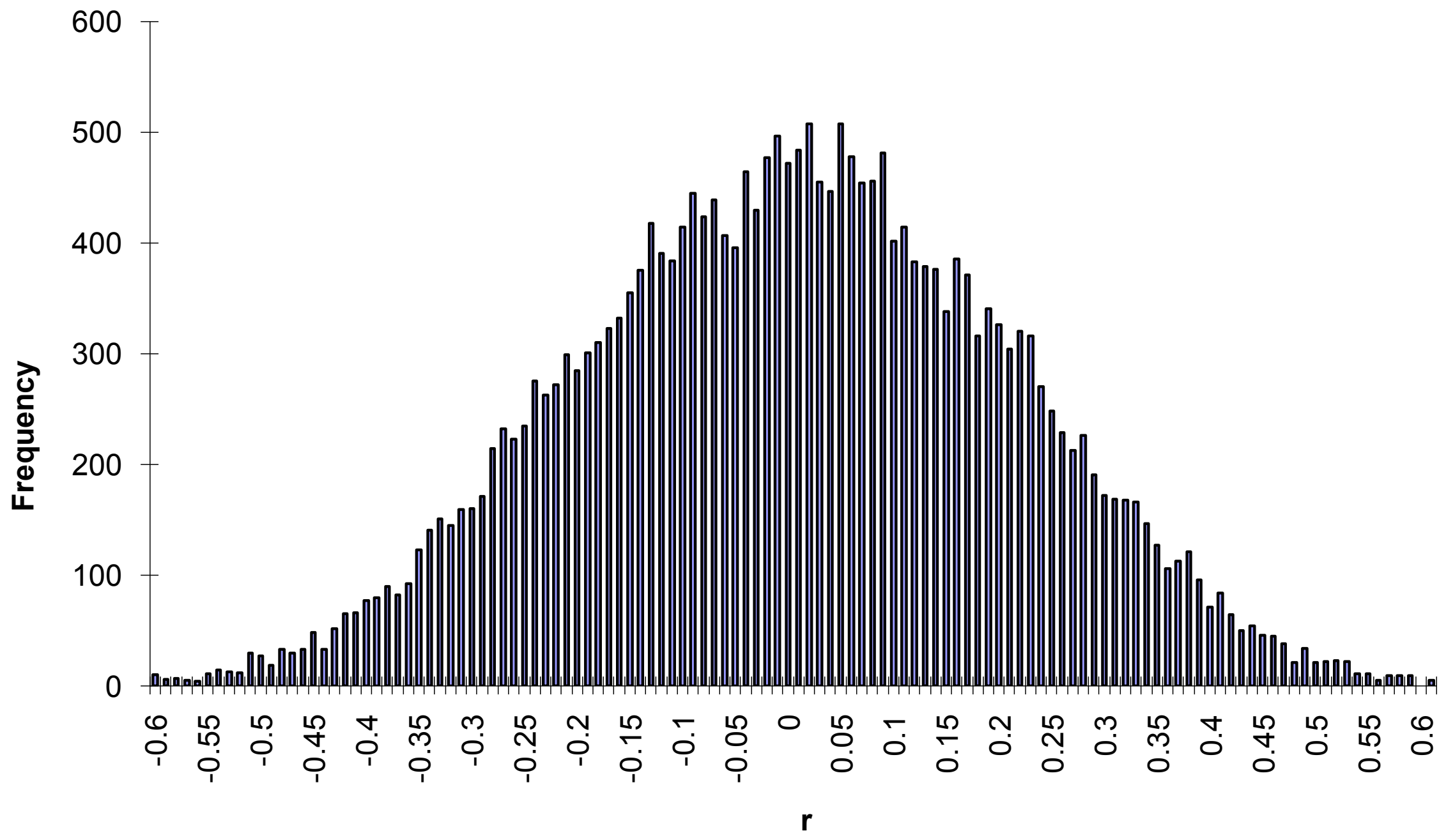

**Figure 13** Correlation between random events and sunspots. A histogram for 25,000 randomly generated events.

In order for their discoveries not to be accidental observations physicists demand confidence levels of 99.99997% (five sigma), or about 1 in 3.5 million chances of being wrong. "Softer" scientists may settle for confidence levels of 99.73% (three sigma), or 27 in 10,000. One way or another, the evidence for the existence of the Kondratieff cycle presented in this article would be challenged to claim scientific rigor and thus acquire the characterization of "a scientific discovery". Therefore economists like Murray Rothbard will be allowed to continue denying the very existence of this phenomenon.

**References**

[1] Jevons, William Stanley, Commercial crises and sun-spots, *Nature* xix, (14 November 1878), pp. 33–37.

[2] Henry Ludwell Moore, *Economic Cycles: their Law and Cause*, University of Michigan Library, January 1, 1914; see also Henry Ludwell Moore, *Generating Economic Cycles*, Macmillan, New York, 1923.

[3] N. D. Kondratieff, The Long Wave in Economic Life, *The Review of Economic Statistics*, vol. 17 (1935):105–115; see also Nikolai Kondratieff (Author), Guy Daniels (Translator), Julian M. Snyder (Introduction), *Long Wave Cycle*, E. P. Dutton, Boston, 1984.

[4] J. A. Schumpeter, *Business Cycles*, McGraw-Hill, New York, 1939.

[5] Bert de Groot and Philip Hans Franses (2008). Stability through cycles, *Technological forecasting and social change*, vol. 75, no. 3 (2008): 301-311.
———— Cycles in basic innovations, *Technological Forecasting and Social Change*, vol. 76, no. 8 (2009): 1021-1025.
———— Common socio-economic cycle periods, *Technological Forecasting and Social Change*, vol. 79, no. 1 (2012): 59-68.

[6] Andrey V. Korotayev, Julia Zinkina, and Justislav Bogevolnov, Kondratieff waves in global invention activity (1900–2008), *Technological forecasting and social change*, vol. 78, no. 7 (2011): 1280-1284.

[7] Two recent issues can be found online here:
http://www.sociostudies.org/almanac/k_waves/k_waves_1_en/articles/
http://www.sociostudies.org/almanac/k_waves/k_waves_2_en/articles/

[8] Murray Rothbard, The Kondratieff Cycle Myth, *Inflation Survival Letters* (14 June 1978); see also, The Kondratieff Cycle: Real or Fabricated? *Investment Insights*, (August and September 1984).

[9] Hugh B. Stewart, *Recollecting the Future: A View of Business, Technology, and Innovation in the Next 30 Years*, Dow Jones-Irwin, Homewood, IL, 1989.

[10] Cesare Marchetti, Fifty-Year Pulsation in Human Affairs, Analysis of Some Physical Indicators, *Futures*, vol.17, no. 3 (1986): 376–88.

[11] Theodore Modis, *Predictions: Society's Telltale Signature Reveals the Past and Forecasts the Future*, Simon & Schuster, New York, 1992.

[12] *Historical Statistics of the United States, Colonial Times to 1970*, vols. 1 and 2, Bureau of the Census, Washington DC; see also *Statistical Abstract of the United States*, U.S. Department of Commerce, Bureau of the Census, Washington, DC.

[13] Hugh B. Stewart, *Recollecting the Future: A View of Business, Technology, and Innovation in the Next 30 Years*, Dow Jones-Irwin, Homewood, IL, 1989, pp. 63-64; see also Theodore Modis, *Predictions: Society's Telltale Signature Reveals the Past and Forecasts the Future*, Simon & Schuster, New York, 1992, p. 266.

[14] J. Ausubel, A. Grubler, N. Nakicenovic, "Carbon Dioxide Emissions in a Methane Economy," *Climatic Change*, vol. 12 (1988), p. 254.

[15] Gerhard Mensch, *Stalemate in Technology: Innovations Overcome the Depression*, Ballinger, Cambridge, MA, 1979.

[16] The clustering of chemical element discovery was first pointed out by Cesare Marchetti, Stable Rules in Social and Economic Behavior, *Bollettino degli Ingegneri,* vol. XXXVI (6), 1988, pp. 3-14.

[17] Theodore Modis, *An S-shaped Adventure: Predictions – 20 Years Later*, Growth Dynamics, Lugano, Switzerland, 2014.

[18] Such a graph was first published with data up to 1980 by Nebojsa Nakicenovic, "Dynamics of Change and Long Waves," report WP-88-074, June 1988, International Institute of Applied Systems Analysis, Laxenburg, Austria. He smoothed the index over a rolling 25-year period with respect to a 50-year moving average.

[19] Gerald S. Hawkins, *Stonehenge Decoded*, Souvenir Press, London, 1966.

[20] Tables for the calculation of tides by means of harmonic constants, International Hydrographic Bureau, Monaco, 1926.

[21] Kenneth C. Land, S-Curves Everywhere, *Science,* vol. 259, 26 February 1993, pp. 1349-1350.

[22] Leonid E. Grinin, Anton L. Grinin, and Andrey Korotayev, Forthcoming Kondratieff Wave, Cybernetic Revolution, and Global Ageing, *Technological Forecasting & Social Change*, vol. 115, pp. 52-68.

[23] Theodore Modis, Sunspots, GDP, and the Stock Market, *Technological Forecasting & Social Change*, 74, 2007, pp. 1508-1514

**Endnote**

[1] Theodore Modis is a physicist, strategist, futurist, and international consultant. He is author/co-author to over one hundred articles in scientific and business journals and eight books. He has taught at Columbia University, the University of Geneva, at business schools INSEAD and IMD, and at the leadership school DUXX, in Monterrey, Mexico. He is the founder of Growth Dynamics, an organization specializing in strategic forecasting and management consulting: http://www.growth-dynamics.com